# Beaming of helical light from plasmonic vortices via adiabatically tapered nanotip

*Denis Garoli[†1], Pierfrancesco Zilio[†1], Yuri Gorodetski[†2*], Francesco Tantussi[1] and Francesco De Angelis[1]*

[1] Istituto Italiano di Tecnologia, via Morego 30, I-16163, Genova, Italy.

[2] Mechanical engineering department and Electrical engineering department, Ariel University, Ariel, 40700, Israel

† The authors contributed equally to the present work

* Corresponding author: Dr. Yuri Gorodetski, yurig@ariel.ac.il

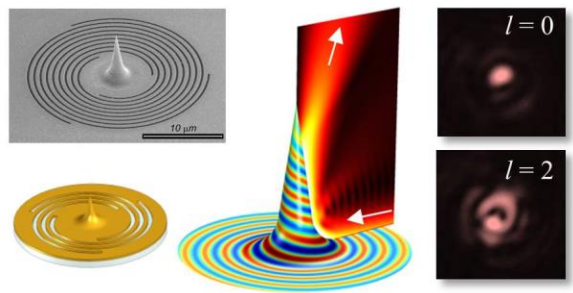

## Abstract

We demonstrate the generation of far-field propagating optical beams with a desired orbital angular momentum by using a smooth optical mode transformation between a plasmonic vortex and free space Laguerre-Gaussian modes. This is obtained by means of an adiabatically tapered gold tip surrounded by a spiral slit. The proposed physical model, backed up by the numerical study, brings about an optimized structure which is fabricated by using highly reproducible secondary electron lithography technique. Optical measurements of the structure excellently agree with the theoretically predicted far-field distributions. This architecture provides a unique platform for a localized excitation of plasmonic vortices followed by its beaming.

Structured light beams have been the subject of an intense work in the last years[1,2] due to the numerous potential applications they may offer in several disparate technological and research fields, ranging from super-resolution imaging[3] to optical tweezing[4] and nanomanipulation[5] to telecommunications.[6]

The possibility to produce and analyze singular optical beams at the micro and the nanoscale led to focus on the interaction of light with metallic nanostructures, resulting in Surface Plasmon Polaritons (SPPs) carrying angular momentum (AM).[7–13] We will refer to these waves as Plasmonic Vortices (PVs). Such modes are generally surface confined helical electromagnetic distributions with a field singularity. The strength of the singularity, termed the topological charge of a vortex, is defined by the phase ramp acquired in one round trip about the singularity center. This charge is proportional to the AM carried by the field.[10]

PVs can be generated by coupling AM-carrying beams to the plasmonic modes of metallic films using particular chiral grating couplers, which have been sometimes called plasmonic vortex lenses (PVLs).[10] Several examples of these couplers have been presented so far.[8–18] A feature common to most of them is the Archimede's spiral shaped grooves or slits milled in a noble metal film.

As has been pointed out in several papers, PVLs can be used not only to couple light to plasmonic vortices, but also to produce strongly localized sources of light carrying non-zero angular momentum.[9,15,16,19,20] In this case, incident circularly polarized light interacting with the plasmonic lens excites PV which is finally scattered into a free space by a proper decoupling structure.

One of the most studied architectures for this aim consists of a PVL with a hole at its center.[9,16,20] The drawback of this scheme is a low efficiency and poor directionality of the far-field beaming due to a small hole size. Moreover, it unavoidably transmits part of the light directly impinging onto the hole, which has not been phase-structured by the PVL. More complex structures, comprising of suspended gold membranes patterned on both sides or multilayer metal-insulator-metal waveguides were proposed to address those issues.[16,20–22] However, besides increasing the fabrication complexity, the highly desirable properties of transmitting a well-defined phase-structured beam with high efficiency and good directionality have been not fully achieved at the same time.

Here we theoretically and experimentally demonstrate a different approach to the efficient PV coupling to the free space, by means of a single-layer PVL structure with a smoothed-cone tip at its center (**Figure 1(a)**). We show that, by properly shaping the tip geometry, the PV excited by the spiral structure can be adiabatically coupled to the far-field mode, carrying well defined AM. A crucial role is played by the strong smoothness of the tip basis. As we shall demonstrate, by simply increasing the curvature radius at the basis of a conical structure, a totally different optical behavior can be achieved, enabling the smooth propagation of the PVs along the tip with a gradual matching to a free space beam propagating normally to the PVL plane. This mechanism is highly efficient for a large range of AM values carried by the impinging PV, with the only losses attributed to the ohmic dissipation. In addition our calculations show that by optimizing the geometrical parameters the circular polarization mixing in the outgoing wave can be kept lower than 15%. As a result, output beams can be theoretically generated at the wavelength scale in an almost pure electromagnetic eigenstate. Interestingly we also demonstrate that, by reciprocity, the structure can be exploited in the reverse way as an efficient localized Laguerre-Gaussian-beam–to-plasmonic-vortex coupler, with back reflected power lower than 1%.

We notice that a number of conical nanotips have already been proposed in combination of PVL architectures,[23–26] but mainly with the purpose of concentrating the axially symmetric $l = 0$ PV in very localized volumes (nanofocusing[27]). For this purpose, a large curvature radius at the tip basis is not needed, or is even detrimental.[26] Other vertical or horizontal rod-like solutions have been proposed with the aim to decouple plasmonic vortexes to the far field,[28–30] Most of the studied conical structures consider an abrupt connection between the antenna basement and the underlying metal surface. In these architectures the decoupling takes place by means of the excitation of one of the antenna electric resonances,[28] whose scattered field pattern is then collected in the far field. The scattering approach to decouple PV has however severe limitations. First, the mechanism can be exploited only to decouple axially symmetric PVs, which have a field maximum in correspondence of the antenna location. PVs with higher topological charge have a field minimum at the PVL center, thus the antenna resonances cannot be excited. Moreover, PVs can hardly be efficiently decoupled out of the PVL surface, most of the power being scattered back again in the form of a PV.

The practical exploitation of our Plasmonic-Vortex-to-free-space mode matcher, presented here for the first time to the best of our knowledge, requires a fine control of the shape of such a non-trivial three-dimensional structure, which is mandatory in order to obtain the desired effect. Here we adopted a powerful technique, sometimes termed as *secondary electron lithography*.[31–33] This fabricative procedure proved to enable a full control of the shape at the nano and microscale, even for exotic 3D nanostructures.[32] Thus a well reproducible nanofabrication procedure allowed fabricating several PVL+tip structures with excellent matching to the designed profile.

In **Figure 1(a)** we show a 3D scheme of the proposed structure. A multiple-turn spiral slit is milled on a 150 nm gold film deposited onto a 100 nm $Si_3N_4$ membrane. Its spiral shape is defined by

$$R_m(\phi) = R_0 + m \cdot \phi / k_{SPP} \tag{1}$$

Here $\phi$ is the azimuthal angle, ranging from *0* to *2πN*, with *N* being the number of spiral turns, $k_{SPP}$ is the propagation constant of the SPP mode propagating at a flat gold-air interface, *m* is an integer denoting the pitch of the spiral, and $R_0$ is the distance from the center to the nearest point of the groove. To maximize the coupling to the plasmonic vortices we consider a set of *m* spirals, each one rotated by *2π/m* with respect to the adjacent ones, in such a way that the radial distance between two adjacent grooves is $\lambda_{SPP} = 2\pi/k_{SPP}$, thus maximizing the coupling to normally impinging light.

The other geometrical parameters of the spiral (slit width and gold thickness) have been tuned to maximize the coupling efficiency of a circularly polarized plane wave impinging from the $Si_3N_4$ side to the SPP modes of the upper gold-air interface (details are reported in the figure caption). A conical gold tip, whose profile is reported in the inset of **Figure 1(a)** is located exactly at the center of the spiral. Three main parameters define the tip shape, height (*h*), apex angle (*α*), and curvature radii at the basis ($r_c$). The curvature radius at the tip apex is kept fixed to 10 nm throughout the paper. As discussed in the experimental section, the adopted fabrication technique enables to faithfully reproduce the designed shape (**Figure 1(b)**).

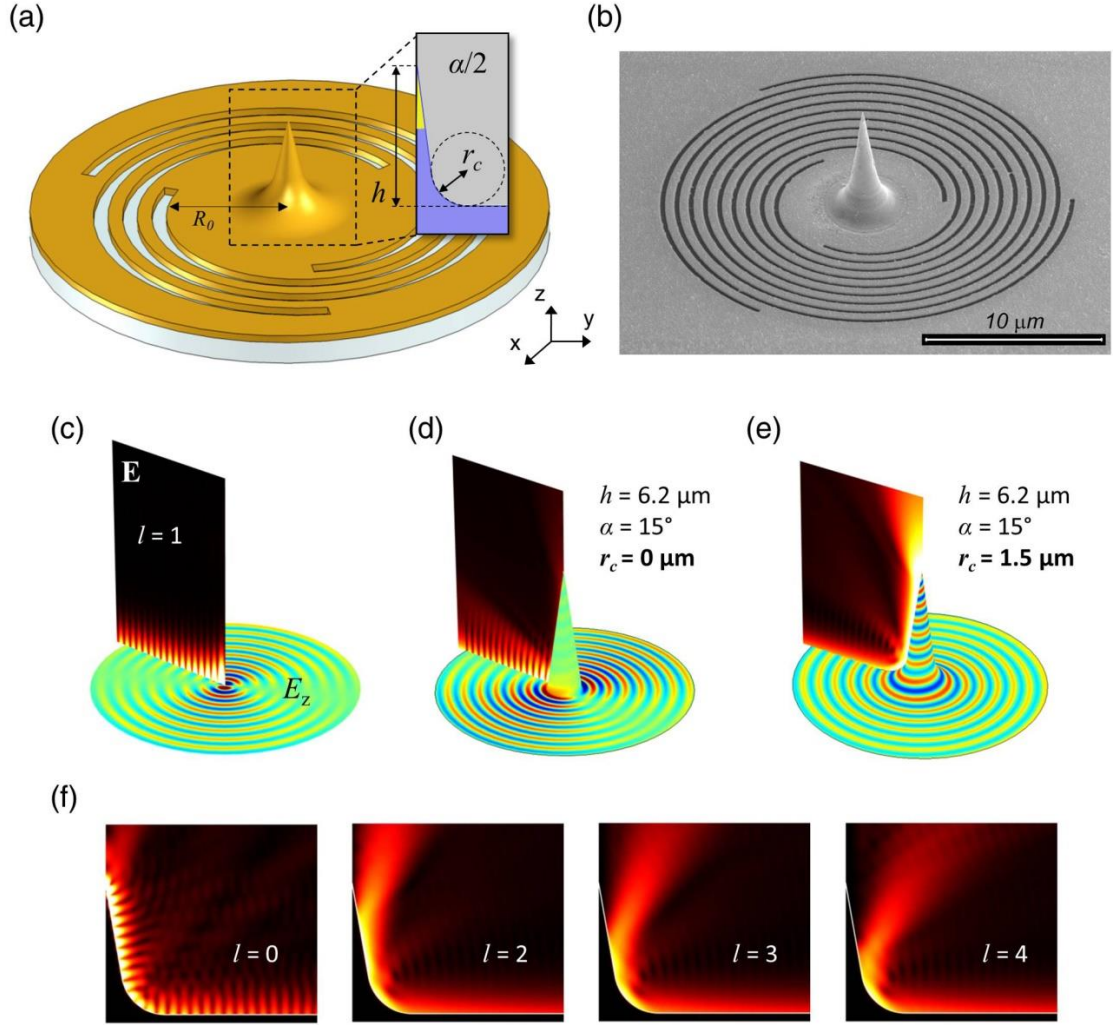

***Figure 1.*** *(**a**) 3D scheme of the PVL. The geometrical parameters of the spiral slits are slit width = 200 nm, pitch = 763 nm, gold thickness = 150 nm, $Si_3N_4$ thickness = 100 nm. $R_0$, namely the initial radius of the spirals, is fixed to 4 µm in order to reduce the effect of direct scattering of light from the grooves. (**b**) Scanning electron microscope image of an example of fabricated structure. (**c-e**) |**E**| maps (vertical cross section) and $E_z$ maps (metal surface plots) of the central region of the PVLs in case of a PV impinging with topological charge l = 1 propagating through a flat PVL center (**c**), scattered by a conical tip with zero basis curvature radius ($r_c$) (**d**), and by a conical tip with large $r_c$ (**e**). (**f**) |**E**| maps in case of a smoothed conical tip (same parameters as in (**e**)) illuminated by a PV with the indicated l values.*

We consider a circularly polarized plane wave impinging normally from the $Si_3N_4$ side. As has been described in several papers,[8–10,14-16, 19] the component of the impinging light electric field locally orthogonal to the slits efficiently couples to the SPP mode of the gold-air interface. The plasmonic field launched by each concentric spiral constructively interferes (thanks to the choice of the spiral period, $\lambda_{SPP}$) producing a PV which radially propagates towards the PVL center. In absence of a tip, the PV confined by the spiral grooves forms a standing wave, giving rise to the characteristic Bessel interference pattern. We use the finite elements software COMSOL Multiphysics® to simulate the electromagnetic distribution in such a flat PVL (see **Figure 1(c)**). The $z$ component of the electric field can be expressed analytically, in cylindrical coordinates, as[10]

$$E_{z,l}(r,\phi,z) = AJ_l(k_{SPP}r)\exp(-\kappa z)\exp(il\phi) \qquad (2)$$

where $k_{SPP}$ is the wave vector of an SPP propagating on a flat gold-air interface, $J_l$ is the Bessel function of first kind and $\kappa = \sqrt{k_{SPP}^2 - k_0^2}$, where $k_0 = 2\pi/\lambda_0$ is the vacuum wave vector. It can be shown[8–10,14] that the topological charge of a PV is given by the relation $l = m + s_i$ where the spin number $s_i = 1$ corresponds to the right hand and $s_i = -1$ to the left hand circularly polarized light.

When the conical tip is present at the PVL center the surface confined electromagnetic mode may couple to the guided mode propagating along the tip upwards. In **Figure 1(d, e)** we compare the simulations of the same PV as in **Figure 1(c)** converging towards two structures having similar conical tips (height $h = 6200$ nm and an apex angle $\alpha = 15°$) but with different curvature radii at the basis. In the case of the negligible curvature radius ($r_c \to 0$) most of the energy is reflected back, which can be clearly deduced from the strong fringe modulation of the intensity (see **Figure 1(d)**). This resembles common configurations presented elsewhere.[23–26] The fraction of the PV power incoming at the tip basis scattered to the far-field (we term this quantity transmittance hereafter) in this configuration is rather low (less than 10%).

This behavior dramatically changes with increasing of the curvature radius at the tip basis. This is shown in **Figure 1(e)**, where we consider the tip basis curvature of $r_c = 1.57$ μm. By looking at the field maps, we now see a smooth intensity pattern close to the metal surface around the tip, which means the absence of back reflected SPP at the tip basis. This is also demonstrated by the $E_z$ field pattern on the metal surface, which shows perfectly spiraling wave fronts propagating along the tip. The smoothed tip in this configuration perfectly matches the PV with the corresponding plasmonic mode of the conical waveguide. Most of the PV power is finally delivered to the free space as a z-oriented beam.

**Figure 1(f)** reports the /**E**/ maps for the PVs with $l$ = 0, 2, 3, 4. As can be seen, for $l > 1$, the behavior is similar and the PV are efficiently decoupled as doughnut shaped waves propagating in the free space. The case of $l = 0$ is an interesting exception. This PV propagates to the end of the tip and is almost fully reflected back, which can be deduced by observing the interference pattern along the metal surface.

A study of the PV transmittance as a function of $l$ for different values of $r_c$ is summarized in the **Figure 2(a)**. It is clearly seen that the increasing of $r_c$ leads to a gradual improvement of the out-coupling efficiencies of the PVs, which can reach values as high as 90%, even for large $l$s. The only losses can be ascribed to the metal absorption. We notice, however, that the coupling of the PV with $l = 0$ to the far field is *always* very poor.

In order to understand the PVs propagation along the cone and their final decoupling we consider the modes of a conducting cylindrical waveguide placed along z-axis, whose radius, $\rho$, progressively decreases with $z$. Their electric field can be expressed as [34,35]

$$\mathbf{E}_i(\rho, z) = \tilde{\mathbf{E}}_i(k_{r,i}\rho)\exp(i\beta z)\exp(il\phi) \tag{3}$$

where the subscript $i = 1,2$ denotes the region outside and inside the cylinder, respectively, $\beta$ is the complex propagation constant of the mode and $k_{r,i}$ is the transverse wave vector, such that $\varepsilon_i k_0^2 = \beta^2 + k_{r,i}^2$, being $k_0 = \omega/c$ the vacuum wave vector and $\varepsilon_1 = 1$, $\varepsilon_2 = -24.1+1.7i$ the relative permittivities of vacuum and gold.[36] The mode amplitude $\tilde{\mathbf{E}}_i(k_{r,i}\rho)$ as well as $\beta$ can be obtained from solution of the Helmholtz equation in cylindrical coordinates in the metal and air domains respectively via imposing the continuity of the tangential components of **E** and **H** fields at the metal surface. This procedure yields a well-known dispersion equation[35] which we solved numerically. Details of the real and imaginary parts of the modes effective index as a function of the cylinder radius are given in the Supporting Information.

As extensively discussed in literature[35] the mode with $l = 0$ has a diverging mode index and experiences zero group velocity close to the $\rho = 0$. This is the reason why such conical structures have been widely utilized for nano-focusing purposes.[26,27] In case of adiabatic tapering, corresponding to the small variations of the plasmon wave number on the scale of a plasmon wavelength, the mode progressively slows down to an almost full stop, leading to a giant concentration of energy at the nanoscale volume.[26] In this case a zero transmittance to the far field would be expected. In our case, the tip aperture is small but non negligible, so the abrupt termination results in reflections from the tip end, clearly visible in **Figure 1(f)**, and in a partial scattering to the far field. The latter is the cause of the low transmittance calculated for $l = 0$ in **Figure 2(a).**

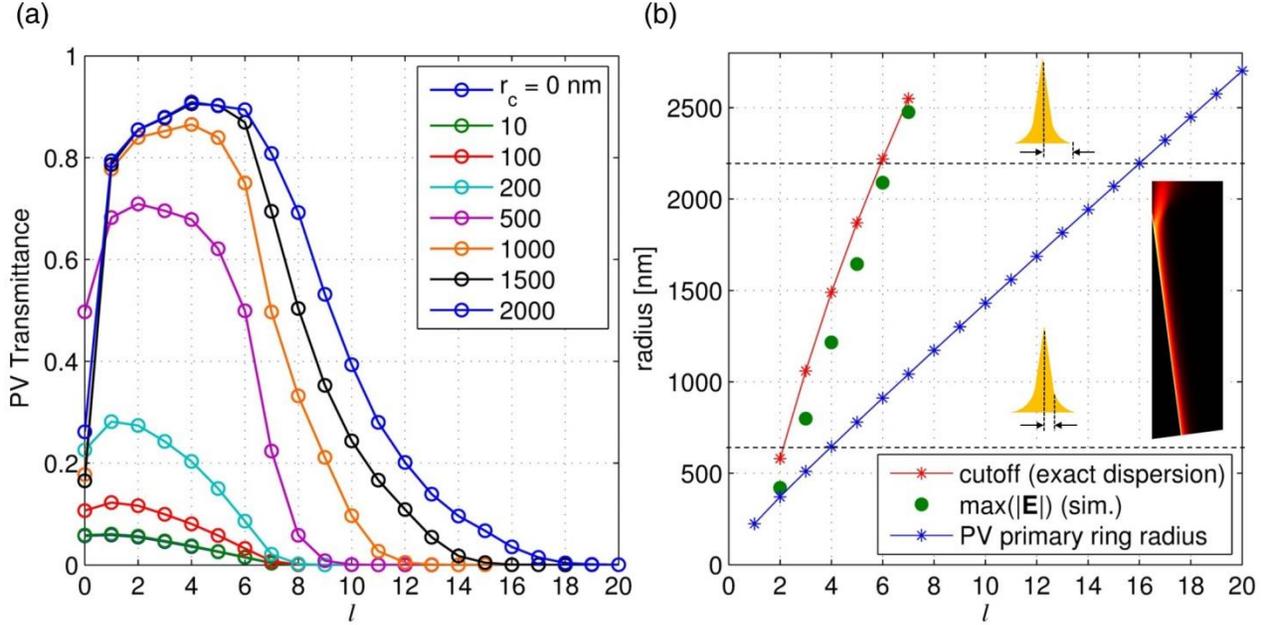

*Figure 2.* *(a) Fraction of the PV light power transmitted to the air domain as a function of the PV topological charge, l, for increasing values of $r_c$. (b) Cut-off radii of the plasmonic modes propagating along the tip calculated by solution of the exact cylindrical mode dispersion equation (red) compared with those ones extracted from the finite elements simulation (green circles); the blue asterisks denotes the primary ring radii of PVs propagating on a flat gold-air surface; the horizontal dashed lines mark respectively the inner tip radius before the smoothing, 650 nm, and the maximum tip radius, 2200 nm; the inset shows an example of electric field norm map calculated in a model of semi-infinite gold tip.*

The $l = 1$ mode also needs a separate discussion. In this case the group velocity adiabatically approaches $c$ and its propagation constant tends to $k_0$ as $\rho \to 0$.[35] This mode is, therefore, guided to the tip end, where it efficiently decouples to the free space. The perfect momentum matching in this case assures no back reflections and most of the energy is beamed out. Apparently with this mode one obtains a Gaussian-beam- like source of radiation, rather than a localization of the energy at the tip end, which is important in applications where high optical throughput is needed.

All other modes, $l > 1$, exist in a bound form only for $\rho$ larger than some $l$-dependent cut-off value, at which the modal loss vanishes, $\text{Im}(\beta) = 0$.[37] To verify that such a mode conversion mechanisms takes place along our conical tip, which has a small but non-negligible aperture $\alpha = 15°$, we simulate just a portion of a very long metal tip, exciting the *l*-th mode at the basis boundary (**Figure 2(b)**, inset). This enables to exclude from the analysis the effects of the finite length of the tip, thus allowing studying arbitrarily high *l* values. We empirically estimate the modes detaching points by considering the points at which the electric field norm along the metal surface reaches its maximum. The corresponding radii for *l* from 1 to 7 are marked in **Figure 2(b)** with green dots. As can be seen, the cutoff values obtained from solution of the exact dispersion equation (red asterisks) are in good agreement with the numerical simulation. We notice that the tip we consider ($h = 6200$ nm, $r_c = 1570$ nm) has a maximum radius of about 2.2 µm, at the beginning of the smoothed part, while the tip radius at the fillet end is just 650 nm (these radii are marked with horizontal dashed lines in **Figure 2(b)**). This latter is the radius at which the tip slope effectively starts to be equal to $\alpha/2 = 7.5°$. This radius is just larger than the cutoff of $l = 2$ mode. Nonetheless, a high transmittance is predicted for all modes up to $l = 6$, as can be seen in **Figure 2(a)**, black line. For these modes the large curvature radius at the basis ensures a smooth transition between the plasmon modes and the free propagating waves in air. For $l > 6$ transmittance rapidly drops (**Figure 2(a),** yellow line) becoming zero for $l > 15$. As a matter of fact, we notice that when $l > 6$ the cutoff radius is larger than the maximum tip radius. For *l* up to 16, however, we observe that the first maximum of the Bessel interference pattern that would arise in absence of tip (sometimes called the primary ring,[9] blue line in **Figure 2(b)**) occurs at a radius smaller than the maximum tip radius. This enables an interaction with the tip basis which determines a partial decoupling of the PV to the free space. For $l > 16$, instead, the primary ring radius is larger than the maximum tip radius, and therefore the PV does not interact with the tip at all, resulting in a situation similar to a flat PVL like in **Figure 1(c).** Accordingly, in this situation the decoupled power is zero. In order to increase the maximum *l*-th mode interacting with the tip, an higher tip with the same curvature radius at the basis can be prepared (as can be evinced from **Figure 2(b)**).

Once the *l*-th mode detaches from the tip it propagates in air as a wave with the unique spatial phase structuring. The radial polarization of the PV at the tip surface collapses in the free space into the scalar components of two opposite circular polarizations, $\mathbf{E}_\pm = \mathbf{E} - is_o\mathbf{Z_0H}$ [20] carrying OAM of $l_o = l - s_o = l \mp 1$, where the OAM of the PV is $l = m + s_i = m \pm 1$; $s_i$ is the incident spin number and $s_o$ is the spin of the outcoupled light. We can summarize the expected OAM content of the output cross circular polarization components in the following table: [20]

***Table 1.*** *Topological charges of the outcoupled beam depending on the incident and the emerging polarization state.*

| Table 1 | | |
|---|---|---|
|  | $s_i = +1$ | $s_i = -1$ |
| $s_o = +1$ | $m$ | $m-2$ |
| $s_o = -1$ | $m+2$ | $m$ |

As has been stressed elsewhere,[38] the process of the impinging light scattering results in the appearance of some amount of a spin-flip component, that is the light of opposite handedness with respect to the incident one. This mix of the output polarization states can be large and depends, besides the geometry of the structure, on particular symmetries of Maxwell's equations[39,40]. In this regard, we notice that our structure behaves as a *mode matcher* rather than a scatterer. Specifically, the coupling of the incident circularly polarized light to PV followed by its guidance to the tip apex incorporates the 3D transformation of the electromagnetic fields. The surface curvature combined with an effective index change along the tip forces the emerging spin to be $s_0 = sgn(l)$[40]. In other words, for positive topological charges of the PVs the tip emits the light with almost pure spin state $s_o = 1$. The tip profile we present has been optimized in order to simultaneously maximize, for *l* ranging from 1 to 4, both the tip transmittance and the *polarization contrast*. The latter is defined as $Q = 1 - P_+/P_-$, with $P_+$ and $P_-$ being the light powers decoupled by the tip with right and left circular polarization state respectively. A plot of transmittance and Q for the first *l*s is reported in **Figure 3(a)**. As can be seen, Q remains higher than 83% for *l* =1 to 4. For more details about the tip parameters optimization we remand to the Supporting Information.

Finally we want to characterize the propagating waves decoupled by the tip in terms divergence and intensity profile. It is important to underline that, unlike other configurations presented in literature,[28–30] in this case what is transmitted to the far field is a well-defined beam with a single intensity lobe. **Figure 3(b)** reports the electric field norm profiles for *l* = 1 to 5 calculated on a horizontal cross section 1 µm above the tip. We notice that profiles are well fitted by Laguerre-Gaussian beam shapes (dashed black lines), namely

$$LG^0_{l_o}(r) = \frac{a}{w(z)}\left(\frac{r}{w(z)}\right)^{|l_o|} \exp\left(-\frac{r^2}{w_0^2[1+i(z/z_R)]}\right) \quad (4)$$

with $l_o = l - s_o$ the OAM of the outgoing wave, $w(z) = w_0[1+(z/z_R)]^{0.5}$, $z_R = \pi w_0^2/\lambda$ the Rayleigh range and adopting as parameters $a$, $z$ and the beam waist, $w_0$. The inset of **Figure 3(b)** shows the far-field polar plots for each $l$. Clearly the peak intensities are found within a cone of maximum 40° half-aperture. **Figure 3(c)** presents the beams divergences (i. e. the angles from the normal at which the intensity drops to 1/e times the peak value) for $l$ = 1 to 5. We compare the cases of finite-height smoothed tip and the infinite tip are considered. As can be seen, the divergences are similar only for $l$ = 1 and 2, since for these $l$s, in both the realistic and idealized tip models, the PVs propagate along the tapered part of the tip before detaching. Nonetheless, for all $l$ from 2 to 5 the beam divergences are lower than about 60°. Therefore, the output beams are readily measurable by standard microscope objectives, (an angular apertures of 60° corresponds to a numerical apertures of 0.866). Interestingly, these results suggest that the smoothed conical structure can be exploited also as a perfect Laguerre-Gaussian-beam-to-Plasmonic-Vortex converter. As an example, in **Figure 3(d)** we simulate the illumination of the tip by normally impinging circularly polarized focused Laguerre-Gaussian beams with $l_o$ = 0, 1 setting the beam parameters to those ones obtained from the fits of the corresponding outgoing waves produced by PV decoupling. As can be seen, the electric field norm distributions show a very smooth coupling to the conical tip modes and finally to the PV modes. The calculated back reflectances are lower than 1%, while the power delivered to PVs at the tip base is 76% for $l_o$ =0 and 84% for $l_o$ =1.

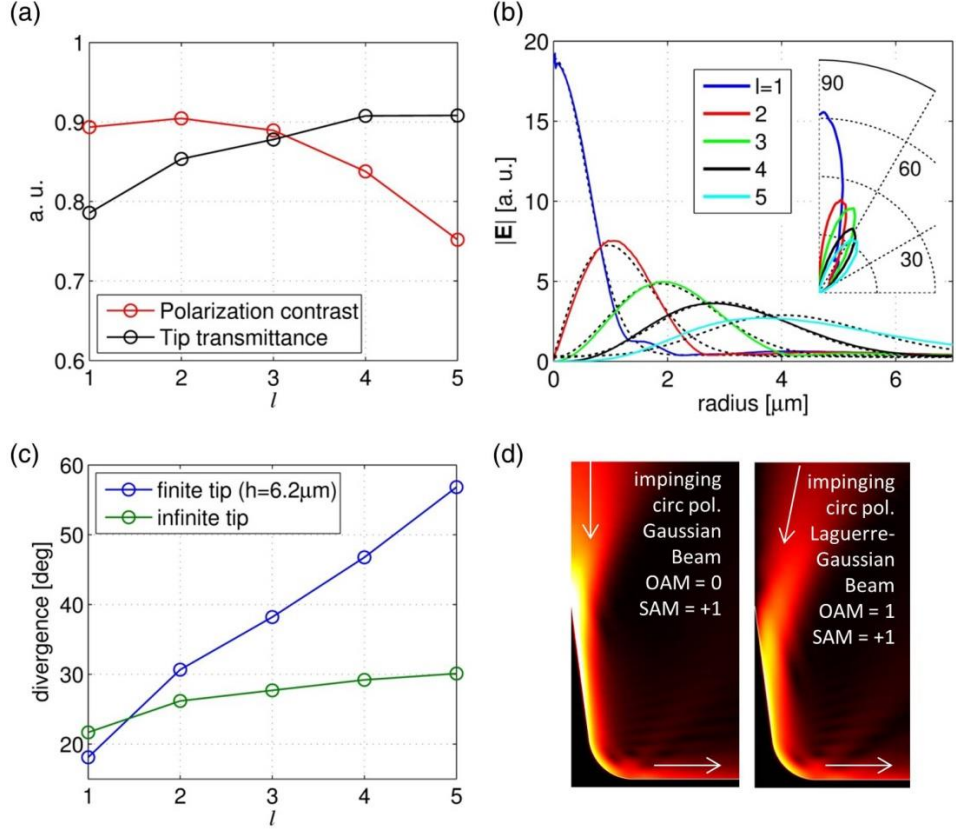

*Figure 3.* *(a) Polarization contrast (red) and tip transmittance (blue) as a function of the PV topological charge, l. (b) Simulated |**E**| field profiles for various l (colored lines) taken at a horizontal cross section 1 μm above the tip, compared with the corresponding fit curves, obtained by using equation (4). (c) Simulated divergences of the beams in case of finite smoothed tip (blue line) compared to the semi-infinite one (green). (d) Examples of simulations of coupling of focussed Laguerre-Gaussian beam to plasmonic vortex, by coaxial illumination of the tip from above. As indicated in the figure, the beams have $l_o = 0$, $s_o = 1$ (Gaussian beam) and $l_o = 1$, $s_o = 1$ (Laguerre Gaussian beam).*

To prove the aforementioned properties of beaming of helical light by means of our adiabatically tapered tip, we provide the experimental demonstrations of the presented analytical and numerical simulations. The sample was illuminated from the bottom with a 20 mW CW single mode pigtailed laser at $\lambda_0$=780 nm. Its spatially filtered and collimated beam was normally incident from the substrate side on the spiral grooves surrounding the tips and an additional optical objective (Olympus

LMPLANFL 100X, NA 0.8) was used to produce an image the tips at CMOS camera (Hamamatsu Orca R2-cooled CCD). According to our simulations, the NA of the imaging objective is large enough to capture the beaming light distribution of up to $l_o = 5$. For higher topological charges, a near-field scanning microscopy might be used due to the strong angular deviation of the scattered light. We used a set of a linear polarizer (LP) followed by a quarter wave plate (QWP) to tune the incident polarization state and an additional set of QWP and a LP to analyze the emerging polarization.

For the sample fabrication an optimized procedure was developed (fully described in Supporting Information). The fabrication of such high tips with arbitrary profile cannot be performed by means of the well-known electron beam induced deposition (EBID)[41] because the process is not stable enough. This limitation was solved by using an approach already used in high aspect ratio 3D plasmonic structures fabrication.[31–33] Briefly, the backbone of the 6.2 µm high tip was prepared, by means of focused ion beam (FIB) (FEI Novalab 600i) exposure on a thick layer of S1813 resist spun on a thin (100 nm) $Si_3N_4$ membrane. A 150 nm gold layer was deposited on the membrane and then the PVL slit patterns with the geometrical vortex topological charges $m$ ranging from 0 to 3 were inscribed on the Au layer by using the FIB. An example of the fabricated nanotip with the surrounding PVL is reported in **Figure 1(b).** A scanning electron microscopy (SEM) image of the PVL with the geometric vortex topological charge of $m = 3$ is illustrated. In the Supporting Information additional examples of fabricated samples can be seen with details on the 3D profile, providing evidence that the obtained tip shape almost perfectly matches the designed one.

**Figure 4** demonstrates the intensity distributions captured by our setup above the structures with $m = 0,1,2,3$. We arrange the results for different polarization states in the order prescribed by the **Table 1**. As expected, our camera captures helical field distributions with topological charge up to $l_o = +5$. Due to unavoidable imperfections the singularity of the beam is split into fundamental first order singularities, providing a convenient way to verify the resultant OAM of the emerging beam, which nicely correspond to the prediction given in the **Table 1**. We note that, as expected, the PVLs with $m = 0$ and $m = 2$ produce point-like emission expressed as an Airy distribution. In some panels we added red arrows to guide the eye to the singular points. For $m = 3$ for $(s_i, s_o) = (+1,-1)$ the beam divergence approaches our imaging limit, therefore the singularities are not clearly visible, however the large primary ring is a clear signature for its high OAM. The helicity conservation effect can be clearly deduced from the PVL structure with $m = 0$. This structure has a circular symmetry, therefore the plasmonic vortex topological charge is $l = s_i$. Accordingly, as stated before, the emerging spin state, $s_0 = sgn(l) = s_i$. We note that in the experiment other sources of helicity change, such as scattering from the spiral rings and direct transmission would limit the polarization contrast to lower values than the ones predicted by our simulations. However, the measured intensity distribution is fully consistent with this selection rule, as one can clearly see the undoubtedly high contrast between the pictures taken with the same polarizer/analyzer state (diagonal) and the ones taken with crossed polarizers (anti-diagonal). Due to specific experiment conditions, the results for higher order PVLs presented in **Figure 4**, do not allow a visual estimation of the polarization contrast, which is one of the topics of our current research.

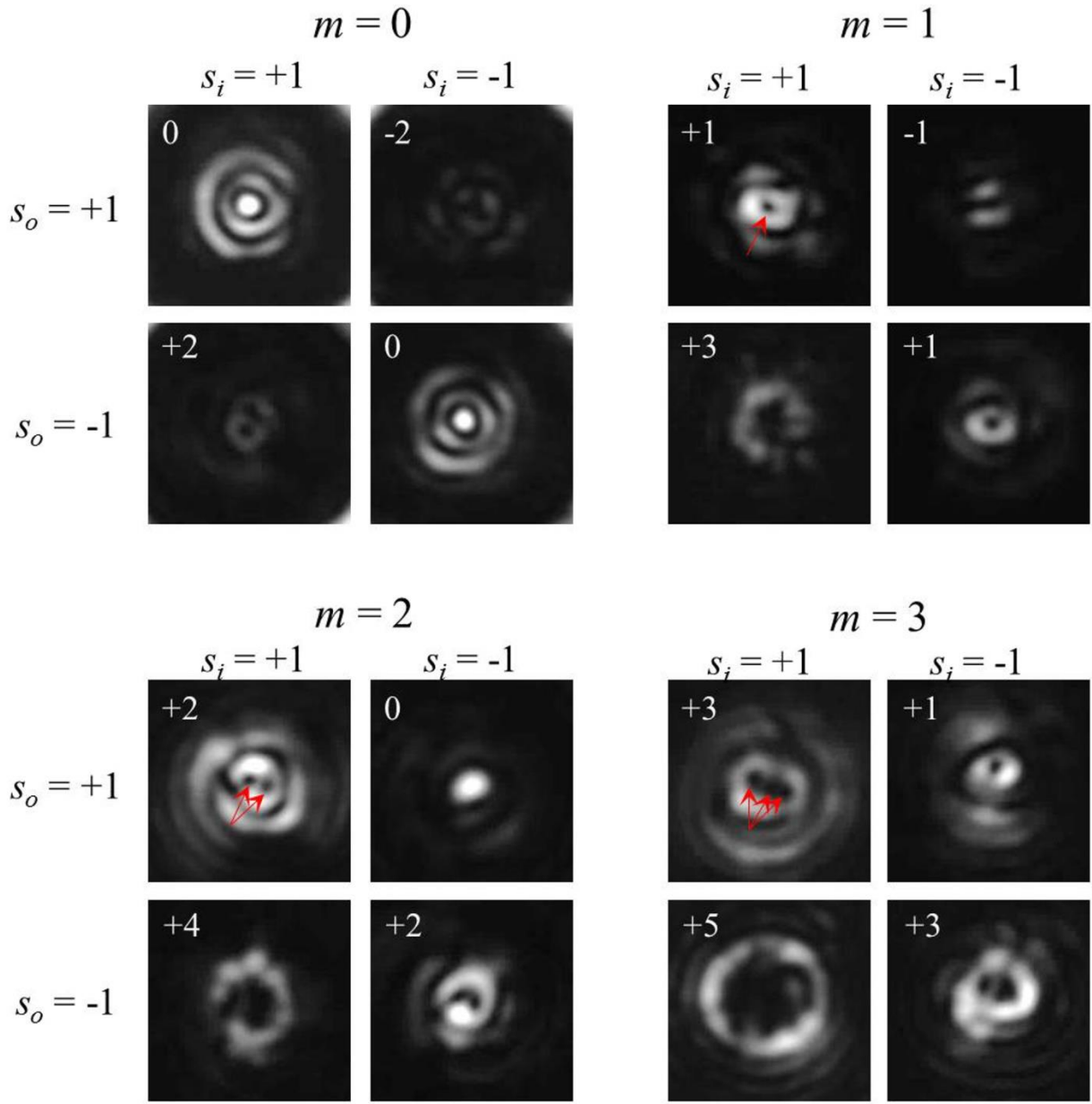

*Figure 4. Far field intensity distributions measured using combinations of circular input polarizers and output analyzers. The images are arranged according to the incident and emerging polarization*

states ($s_i$ and $s_o$, respectively) as in Table 1. The expected topological charges of the far field beams, $l_o$s, are reported in each image and the red arrows are added to guide the eye to the singularity points.

In our last experiment we utilize the fabricated tips as the plasmonic vortex generator in the near-field. In order to measure this we slightly modify the fabrication process by preparing a tip on top of a 60 nm thick gold layer without any grating around it. We illuminate the tip by slightly focused laser beam (20X objective) and image the leakage radiation by using oil immersion 100X objective (NA 1.25). This leakage radiation microscopy system (LRM) provides us a direct image of the plasmonic modes excited at the metal-air interface[42]. **Figure 5** demonstrates the experimental setup along with the measured results. As can be seen from the **Figure 5b**, the stand-alone tip produces plasmonic vortices with $l = \pm 1$, which yields in the far field structured waves with $l_o = 0, \pm 2$, as expected from ideally center symmetric geometry. We verified that the spatial frequency of the measured disturbance, $k = 8.45$ corresponds to the expected plasmonic wavelength. This confirms that our system indeed captures the near field surface waves distribution.

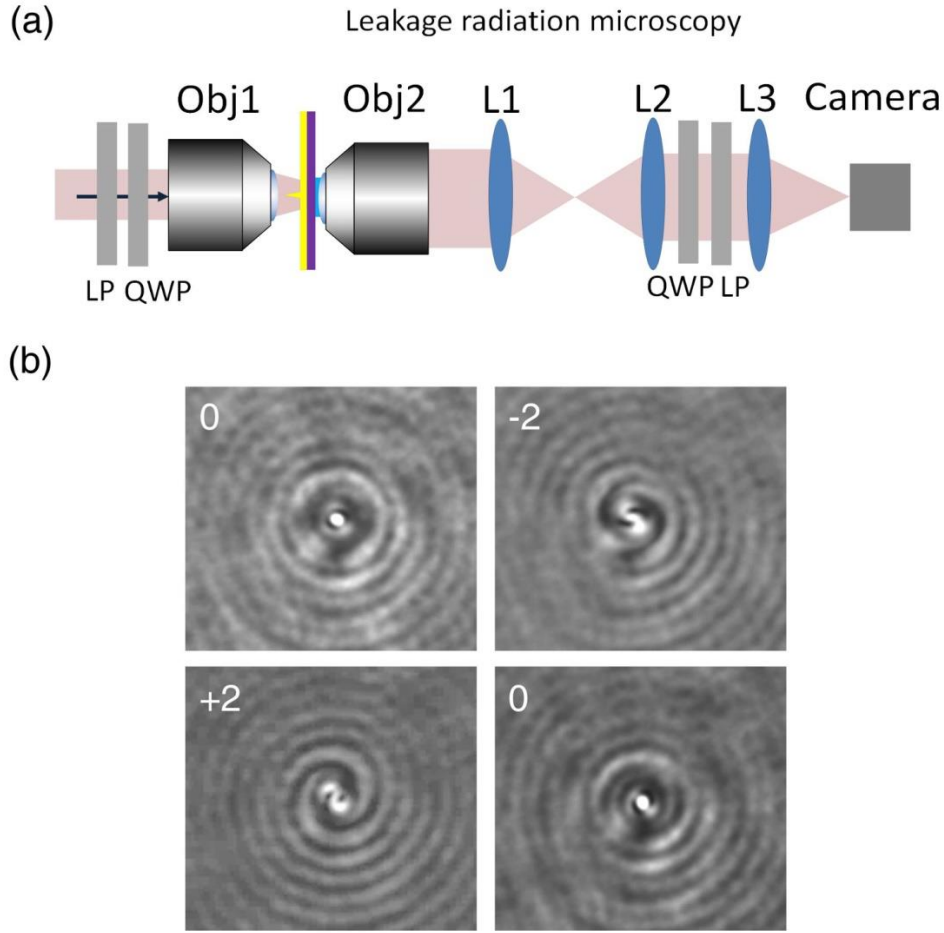

*Figure 5. Near-field PV generation by a nanotip. (a)* – *Scheme of the LRM setup used to measure the near-field. Polarization components* – *LP and QWP were used to tune and analyze the polarization, a 20X objective (Obj1) was used to prefocus the beam on the tip, an oil immersion 1.25NA objective (Obj2) extracted the leakage radiation and a set of the lenses (L1-L3) was used to image the SP distribution. (b) Measured LRM in case of a tip fabricated with 30nm gold coating layer.*

In summary, we presented a plasmonic vortex lens structure able to couple a circularly polarized light to a Plasmonic Vortex and to efficiently transmit it to the far field by means of a smoothed-cone tip placed at its center. The large smoothing introduced at the cone basis is shown to play a crucial role in enabling an adiabatic coupling of the PV propagating on the flat metal surface to the plasmonic modes of the metal tip, whose sharpness, in turn, enables an adiabatic match to the propagating waves in free space. This particular non-trivial geometry has been faithfully experimentally fabricated by secondary electron lithography technique. The optical characterization revealed that phase structured beams were successfully transmitted by the tip up to OAM $l_0 = 5$. Finally, our simulations have shown that this structure can work as an excellent coupler of focused Laguerre-Gaussian beam to PVs. A proof of concept measurement in this illumination condition has been shown, by adopting a leakage radiation microscopy setup.

We believe that such an architecture can be a key element for the realization of compact focused-beams-to-plasmonic-wave couplers, elements which are highly desirable to fully exploit plasmonic technologies in practical applications, such as sensing, light manipulation in flat guided optics, optical tweezing etc. On the other side, a scattering-free, highly directional PV read-out element could be also extremely valuable, enabling to directly visualize any near-field coherent azimuthal interference taking place at the center of a PVL. Thanks to the flexibility of secondary electron lithography (as demonstrated elsewhere[31]), a complete control of the tip shape can be obtained, enabling to finely tune the coupling-decoupling capability of the tip, thus making it interesting for a large number of applications. For example, smaller tips can be used to decouple PVs produced by simple bull's eye structures, while higher tip can enable far field access to arbitrary high-OAM carrying PVs.

This research has been recently published in Nano Letters[43].

## AUTHOR INFORMATION

\* Corresponding author: Dr. Yuri Gorodetski, yurig@ariel.ac.il

## AUTHOR CONTRIBUTION

DG fabricated the structures; PZ ideated, designed and simulated the structures; YG provided theoretical support and performed the optical characterization, FT helped in the optical characterization and optics setup, FDE supervised the work.

## ACKNOWLEDGMENTS

The research leading to these results has received funding from the European Research Council under the European Union's Seventh Framework Program (FP/2007-2013) / ERC Grant Agreement n. [616213], CoG: Neuro-Plasmonics and under the Horizon 2020 Program, FET-Open: PROSEQO, Grant Agreement n. [687089].

# SUPPORTING MATERIAL

# Beaming of helical light from plasmonic vortices via adiabatically tapered nanotip

*Denis Garoli[†1], Pierfrancesco Zilio[†1], Yuri Gorodetski[†2*], Francesco Tantussi[1] and Francesco De Angelis[1]*

[1] Istituto Italiano di Tecnologia, via Morego 30, I-16163, Genova, Italy.

[2] Mechanical engineering department and Electrical engineering department, Ariel University, Ariel, 407000, Israel

† The authors contributed equally to the present work

* Corresponding author: Prof. Yuri Gorodetski, yurig@ariel.ac.il

**Supporting Note 1: Fabrication process of plasmonic vortex lens with central tip.**

The method followed to fabricate the samples is based on a procedure introduced and fully described by De Angelis et al. (see ref. 1). The principle relies on FIB-generated secondary-electron lithography in optical resists and allows the preparation of high aspect ratio structure with any 3D profile. The final structure comprises of a 6.2 μm high base-smoothed gold tip on a 150 nm gold layer where *m*-PVL are milled. In order to prepare such a complex architecture a multi-steps fabrication process have been optimized. First of all a 5 / 23 nm Ti / Au bilayer has been deposited, by means of sputtering, on a 100 nm thick $Si_3N_4$ membrane. On this conductive layer, s1813 optical resist has been spun at 1500 rpm and soft-baked at 90°C for 8 minutes. The resist thickness of 11 μm is achieved by tuning the concentration, spinning time and velocity. On the back of the membrane a thin layer of silver (about 10 nm) is then deposited by means of sputtering in order to ensure the necessary conductibility of the sample for the successive lithographic step. The membranes are then patterned from the backside using a Focused Ion Beam (Helios Nanolab600, FEI company), operated at 30 keV (current aperture: 80pA, dwell time: 500 μs). The tip-like shape has been obtained by patterning successive disks with decreasing diameter and correcting the dose applied for every disk, thus resembling the expected tip profile (**Figure S1(a)**). (To note that the first milled disk present a high thickness (around 80 nm) that will be filled in the successive metallic growth). Due to the high dose of low-energy secondary electrons induced by ion beam / sample interaction, a 30 nm thick layer of resist, surrounding the milled disks, becomes highly cross-linked and insoluble to most solvents. After patterning, the sample is developed in acetone, rinsed in isopropanol and dried under gentle $N_2$ flow. The back side silver layer has been then removed by means of rapid $HNO_3$ rinse. At this stage we get a high dielectric tip surrounded by a metallic substrate. Since we need a base-smoothed tip on a 150 nm thick gold layer, an additional layer of metal has been grown of the substrate by means of galvanic deposition (0.12 Amp

DC). The galvanic layer is grown up to the tip base so ensuring a very smooth geometry (see **Figure S1(b)**). After the galvanic deposition, a 40 nm thick layer of gold is deposited by sputtering the sample, tilted 60° with respect to the vertical and rotated, guaranteeing an isotropic coating on both the sidewalls and the base. (In order to avoid any possible direct transmittance from the tips, the back of them has been filled, by means of electron beam induced deposition, with a 200 nm thick layer of platinum). Finally, in order to prepare the sample with the desired *m*-order PVL surrounding the tip, a FIB milling process has been performed on the sample creating the spiral gratings without affecting the quality of metallic tip. Examples of fabricated PVLs are reported in **Figure S1(c)**.

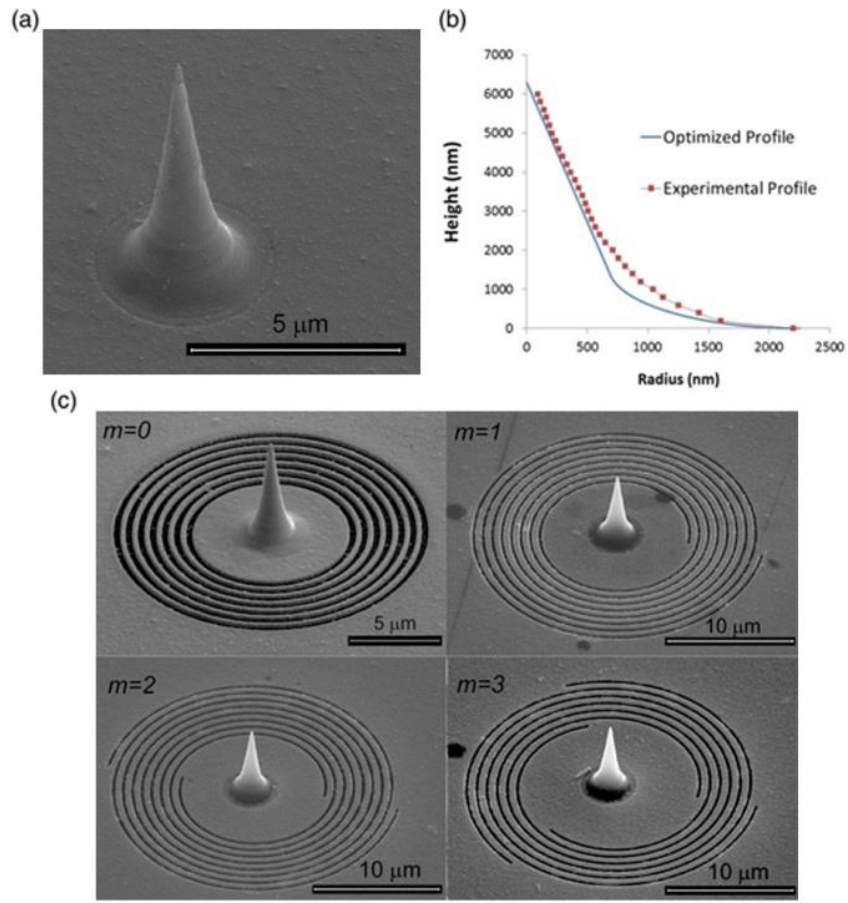

**Figure S1. (a)** SEM image of an isolated tip. **(b)** Comparison between designed and fabricated profile. **(c)** SEM micrographs of PVLs with topological charge ranging from 0 to 4 and tip at the center.

**Supporting note 2: Mode analysis of gold cylindrical waveguides**

We report here the analysis of the guided modes of a gold cylindrical waveguide placed along $z$-axis, as a function of the radius, $\rho$. Their electric field can be expressed as

$$\mathbf{E}_i(r,z) = \tilde{\mathbf{E}}_i(k_{r,i}r)\exp(i\beta z)\exp(il\phi) \tag{S1}$$

where the subscript $i = 1,2$ denotes the region outside and inside the cylinder, respectively, $\beta$ is the complex propagation constant of the mode and $k_{r,i}$ is the transverse wave vector, such that $\varepsilon_i k_0^2 = \beta^2 + k_{r,i}^2$, with $k_0 = \omega/c$ being the vacuum wave vector and $\varepsilon_1 = 1$, $\varepsilon_2 = -24.1+1.7i$ the relative permittivities of vacuum and gold at $\lambda = 780$ nm.[2] The mode amplitude $\tilde{\mathbf{E}}_i(k_{r,i}r)$ as well as $\beta$ can be obtained from the solution of Helmholtz equation in cylindrical coordinates in the metal and air domains respectively via imposing the continuity of the tangential components of **E** and **H** fields at the metal surface. The resulting dispersion equation (reported for example in Ref. 3) was solved numerically. In **Figure S2(a,b)** we report respectively the real and imaginary parts of the mode's effective index ($N_{eff} = \beta/k_0$) at $\lambda = 780$ nm as a function of the cylinder radius for azimuthal numbers ranging from $l = 0$ to 7. The inset of **Figure S2(a)** shows the group velocity, $v_g$, for the first four modes.

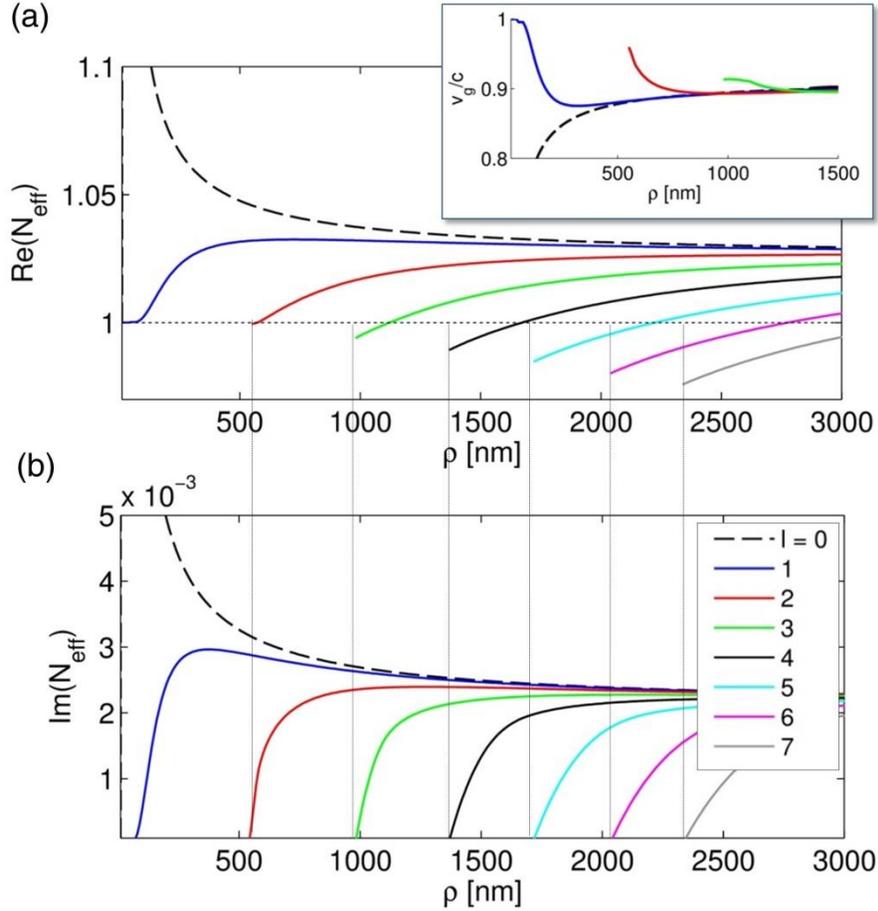

**Figure S2.** Modal analysis of a cylindrical gold waveguide in vacuum as a function of its radius, $\rho$, for azimuthal mode index $l = 0$ to 7 and for a fixed vacuum wavelength $\lambda = 780$ nm. **(a)** Real part of the effective mode index, $N_{eff} = \beta/k_0$; inset: normalized group velocity for the first four modes. **(b)** Imaginary part of $N_{eff}$.

All modes with $l > 1$ exist in a bound form only for $\rho$ larger than some $l$-dependent cut-off value, at which the modal loss vanishes, $\text{Im}(\beta)=0$. As can be seen in **Figure S2(a)**, the modes do not cut off exactly at the momentum matching points, namely at $\beta = k_0$ as usually expected from tapered dielectric waveguides.[4,5] The reason of the different behavior in our case is a consequence of the large imaginary part of the modal refractive index, which causes the transverse index $k_r$ in (S1) to become complex-valued.[4,5]

**Supporting note 3: Tip optimization studies**

In the main text we mention that the tip parameters (curvature radius at the basis, $r_c$, tip aperture, $\alpha$, and tip height, $h$) have been optimized in order to maximize the average tip transmittance (T) and the polarization contrast, Q, for $l$ ranging from 1 to 4. Q is defined as $Q = 1 - P_+/P_-$, with $P_+$ and $P_-$ being the light powers decoupled by the tip with right and left circular polarization state respectively. In **Figure S3(a-c)** we report the behavior of T and, for convenience, $R = P_+/P_-$, as a function of the various parameters for $l = 1$ ot 4. **Figure S3(d)** reports T and R as a function of $l$ for the optimal parameters set.

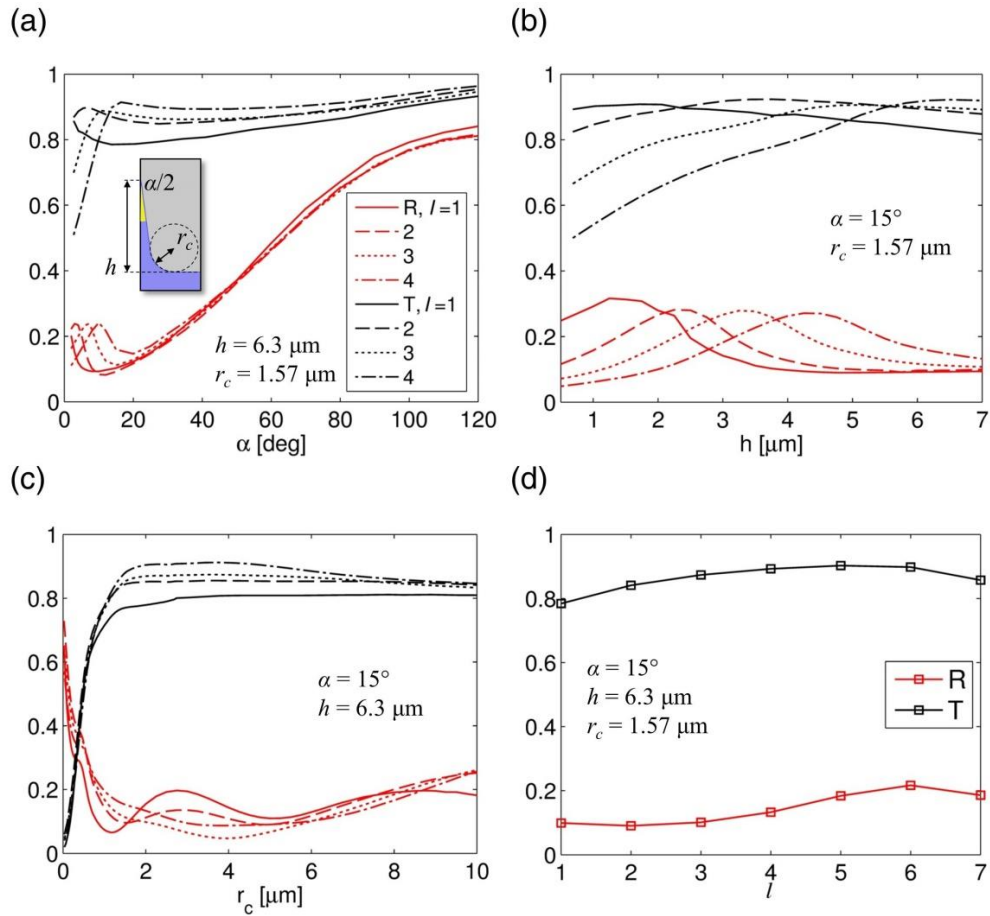

**Figure S3.** Study of tip transmittance, T, and R parameter (details in the text) as a function of the tip parameters, $\alpha$ **(a)**, $h$ **(b)**, $r_c$ **(c)**, for PV topological charges $l = 1$ to 4. For each plot the remaining parameters are fixed to the optimal. **(d)** R and T as a function of $l$ for the optimized tip.

**Supporting References**